# Short Term Reliability and Robustness of ultra-thin barrier, 110 nm-gate AlN/GaN HEMTs

Zhan. Gao[1], Matteo. Meneghini[1], Kathia Harrouche[2], Riad Kabouche[2], Francesca Chiocchetta[1], Etienne Okada[2], Fabiana. Rampazzo[1], Carlo. De Santi[1], Farid Medjdoub[2], Gaudenzio. Meneghesso[1], Enrico. Zanoni[1]
[1]Dipartimento di Ingegneria dell'Informazione, Università di Padova, Via Gradenigo 6/A, 35131 Padova
[2]CNRS-IEMN, Institut d'Electronique, de Microélectronique et de Nanotechnologie, 59652 Villeneuve-d'Ascq, France
Email :gaozhan.veronica@gmail.com

***Abstract*—Short-term reliability and robustness of 110 nm AlN/GaN HEMTs has been evaluated by means of off-state, semi-on state and on-state step stress tests on devices having different gate-drain distance, $L_{GD}$. While breakdown voltages and critical voltages scale almost linearly with $L_{GD}$, failure mode remains almost unchanged in all tested devices, and consists in an increase of gate leakage, accompanied by a positive shift of threshold voltage. In off-state, electroluminescence images detect the presence of localized leakage paths which may act as preferential paths for electron trapping. Degradation does not depend on dissipated power and is preliminary attributed to hot-electron trapping, enhanced by electric fields.**

***Keywords—AlN/GaN HEMTs, reliability, stress, breakdown mechanism, electric field***

## I. Introduction

GaN-based High Electron Mobility Transistors (HEMTs) have drawn great attention due to their potential for high temperature, high frequency and high power applications to radar amplifiers or modern telecommunication systems as 5G [1]. In order to achieve power density and power added effciency at high frequency, up to Q-band, one of the most usual approaches is to scale the gate length of the devices [2], but this would lead to short channel effect. Therefore, it is necessary to optimize the epilayer design and the device layout, and reduce the gate to channel distance and/or adopt wider bandgap materials. like AlN, as barrier [3]. It has been proven that AlN/GaN heterostructures is able to reach a maximum current of 2.3 A/mm and peak transconductance of 480 mS/mm [4], and state-of-the-art PAE over > 65% at 40 GHz [5], by adopting an ultra-thin (3-4 nm) AlN barrier. Good stability and robustness have been achieved by AlN/GaN/AlGaN HEMTs for Ka band operation during constant voltage stress tests [6]. Reliability of AlN/GaN HEMTs with 4 nm AlN barrier during constant voltage stress has been also studied at room temperature and high temperature [7], [8]. However, available data on the reliability and robustness of AlN/GaN HEMTs are still scarce. Effective on-wafer evaluation methods are needed during device development in order to provide a fast feedback on the different technological options.

In this work, we focus on the short-term reliability of a new generation ultra-scaled AlN/GaN HEMTs with 3 nm AlN barrier, and with 110 nm-gate length, grown on SiC substrates. Parametric degradation and breakdown effects observed during drain voltage step stress tests were studied at three states: (1) off-state ($V_{GS}$ = -5 V), (2) semi-on state ($V_{GS}$ = -1 V), and (3) on-state ($V_{GS}$ = 0 V), as well as an off-state constant voltage at ($V_{GS}$ = -5 V, $V_{DS}$ = 40 V) stress test. Degradation and robustness of the HEMTs designed with different gate-drain distance ($L_{GD}$) were compared. Electroluminescence imaging technique was used to have an insight into device behavior during stress. In Section II, the experimental details and the characterization procedures used on the fabricated devices are described. In Section III, parametric degradations during stress, are compared among devices with different $L_{GD}$, the degradation mechanisms during stress are discussed, and the breakdown voltages are summarized at the end of the section.

## II. Experimental Details

Tested structures were fabricated on AlN/GaN heterostructures grown on SiC wafers, with the AlN layer thickness of 3 nm and the GaN channel thickness of 100 nm, grown by MOCVD on top of a carbon doped GaN buffer. In order to reduce surface trapping and better protect the surface, a 10 nm in-situ SiN layer was grown on top, a cross section of the HEMTs is shown in Fig. 1(a). The devices under test are two-fingers devices with gate length ($L_G$) of 0.11 µm, gate width ($W_G$) of 2 × 50 µm; gate-drain distances ($L_{GD}$) are designed to be 0.5 µm, 1.5 µm and 2.5 µm.

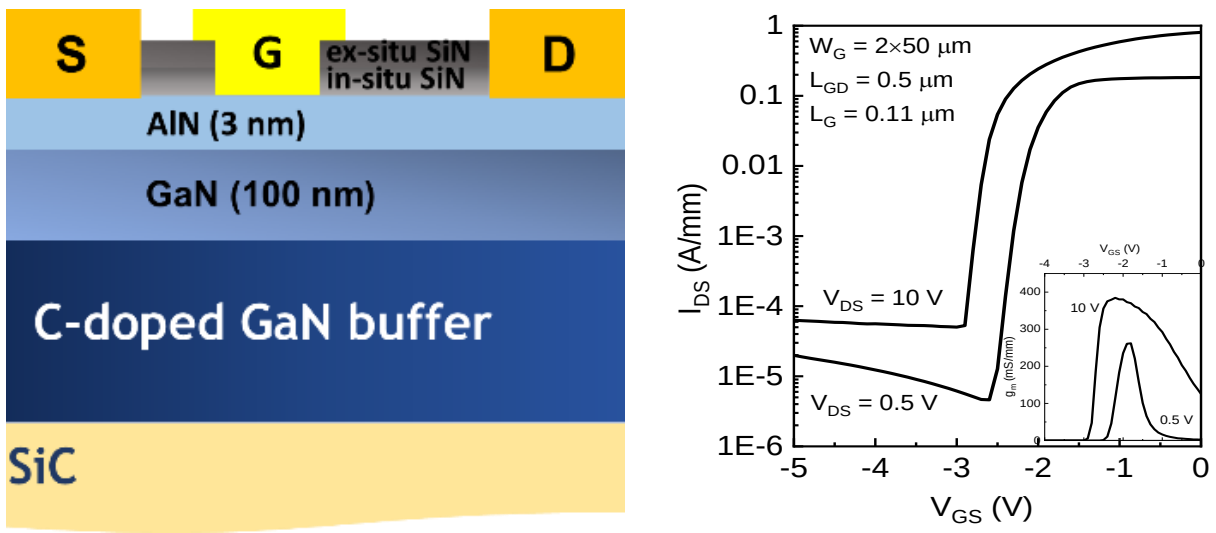

Fig. 1 (a) Schematic cross section and (b) transfer $I_DV_G$, $g_m$-$V_G$ curves of the AlN/GaN HEMTs.

The maximum current of the devices is around 1.2 A/mm and the maximum transconductance is up to 400 mS/mm for the fabricated devices. The transfer $I_D$-$V_G$ and $g_m$-$V_G$ curves of the AlN/GaN HEMTs is shown in Fig. 1 (b). Cut off frequency ($f_T$) and maximum oscillation frequency ($f_{max}$) achieved at $V_{DS}$ = 20V are 63 GHz and 300 GHz, respectively. A more thorough DC and RF characterization is discussed in [5].

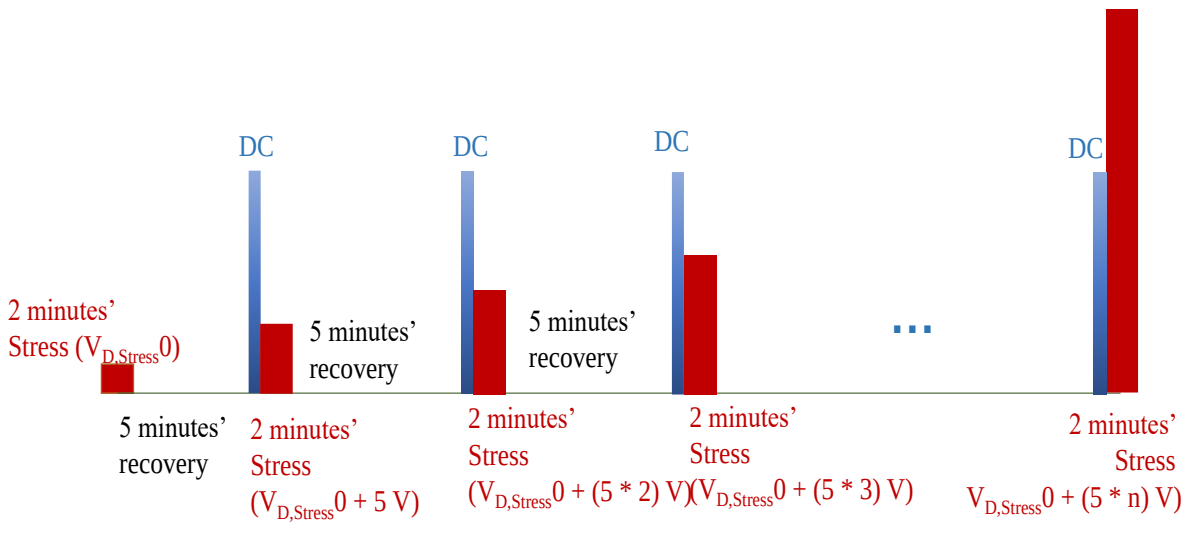

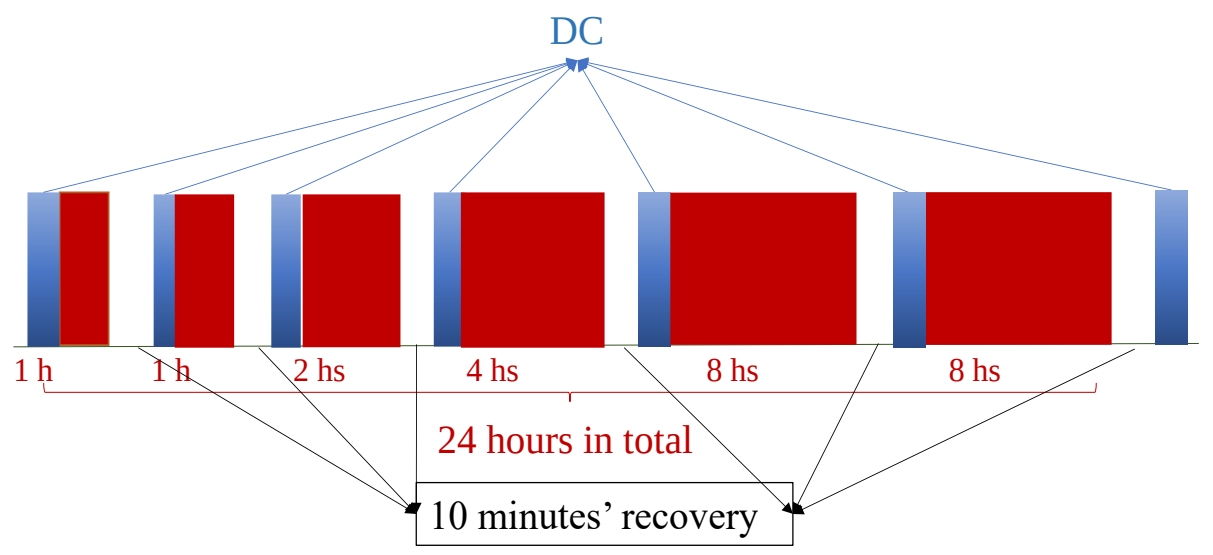

Fig. 2 Flow chart of (a) the step stress and (b) 24 hours' stress

In this work, drain step stress tests at off state ($V_{GS}$ = -5 V), semi-on state ($V_{GS}$ = -1 V) and on-state ($V_{GS}$ = 0 V) were done on devices with different $L_{GD}$. During stress, drain voltage was increased from 0 V to 200 V or up to catastrophic breakdown in 5 V steps, each two minutes long. During each stress step, drain and gate currents as well as EL intensity were monitored.

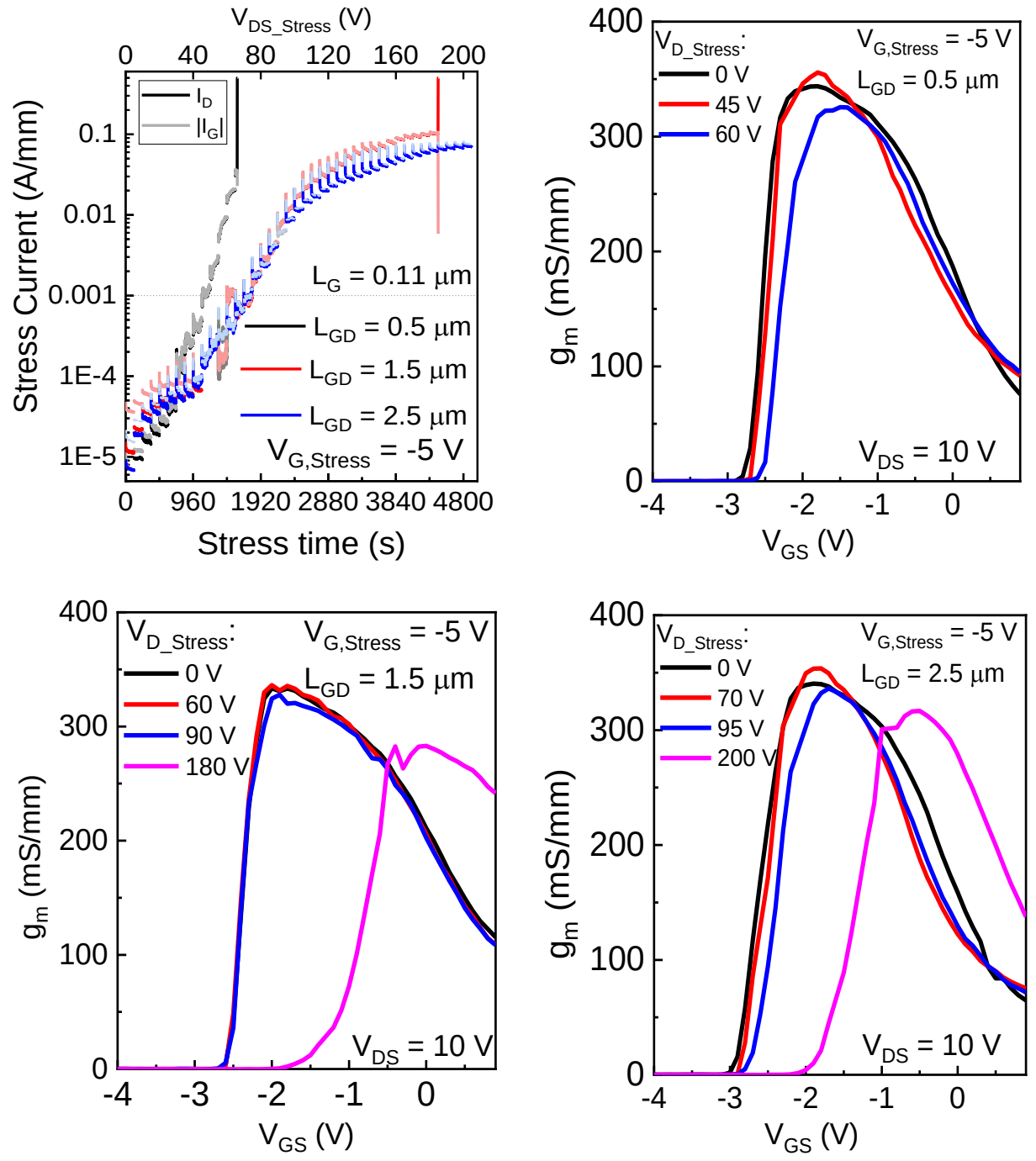

Fig. 3 (a) Drain current monitored during off-state stress and $g_m$-$V_G$ curves during stress (b) $L_{GD}$ = 0.5 µm (c) 1.5 µm and (d) 2.5 µm devices.

After each stress step, the devices were kept unbiased for 5 minutes, and standard DC characterizations were done afterwards, as shown in Fig. 2 (a). A second group of devices was submitted to a constant voltage stress test at off-state ($V_{GS}$ = -5 V, and $V_{DS}$ = 40 V), in order to evaluate the devices reliability on a short time scale. The flow chart of the 24 hours stress is summarized in Fig. 2(b). DC characteristics degradation after the 24 hours' stress among devices having different $L_{GD}$ was summarized and compared.

## III. Results and Discussions

### *A. Off-state drain step stress*

Device drain and gate current recorded during off-state stress ($V_G$ = -5 V) is shown in Fig. 3 (a) for the different $L_{GD}$: 0.5 µm, 1.5 µm and 2.5 µm. Device with $L_{GD}$ = 2.5 µm didn't show catastrophic failure up to 200 V, while $L_{GD}$ = 0.5 µm and 1.5 µm devices suffered destructive breakdown at 70 V and 185 V, respectively. If one adopts a failure criteria corresponding to a drain leakage current density of 1 mA/mm, the following failure voltages are obtained, respectively for $L_{GD}$ = 0.5 µm, 1.5 µm and 2.5 µm devices: 45 V, 60 V and 70 V. The increase in leakage current is usually caused by the generation of preferential leakage current conduction paths.

As the step stress proceeds, a positive threshold voltage shift occurs; in the $L_{GD}$ = 1.5 µm device, $V_{TH}$ shift reaches approximately 1.5 V, at the same time transconductance $g_m$ (at $V_{DS}$ = 10 V) shows a decrease of 16% at the 180 V step. It should be noted, however, that $V_{TH}$ shift is negligible and that no degradation of $I_{DSS}$ or $g_m$ occurs until the stress voltage reaches 50 V, 85 V and 90 V, for $L_{GD}$ = 0.5 µm, 1.5 µm and 2.5 µm devices, respectively.

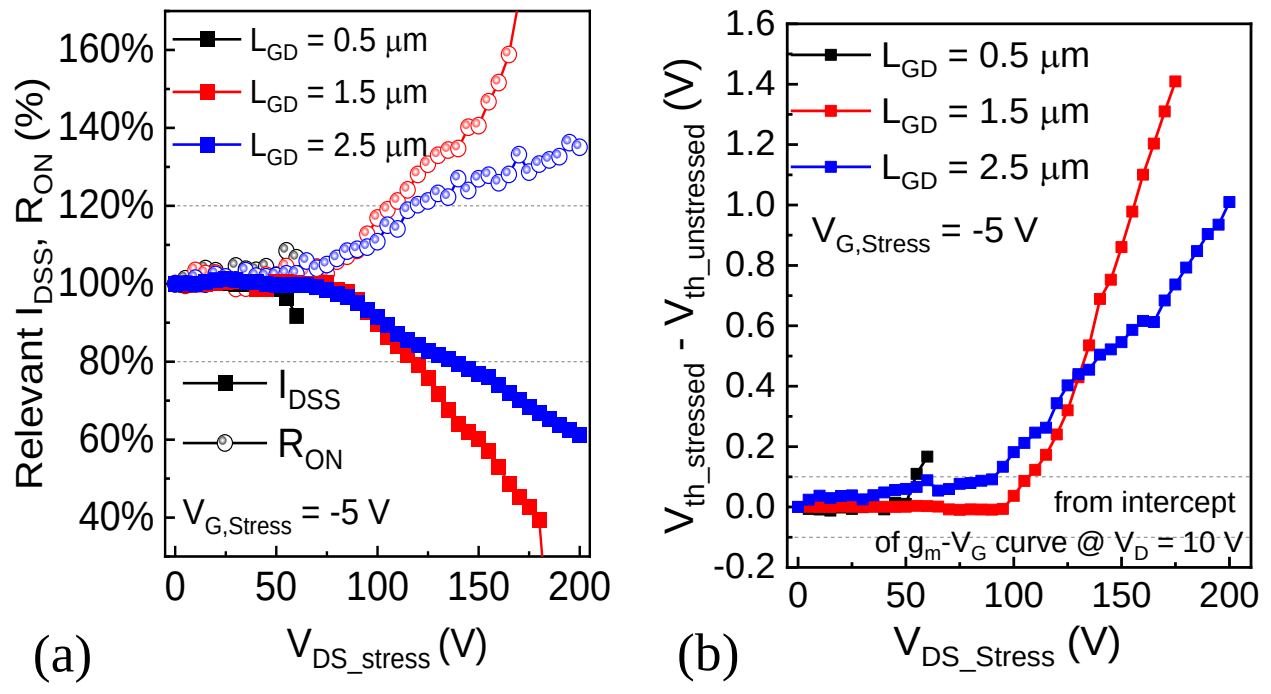

Fig. 4 (a) $I_{DSS}$, $R_{ON}$ and (b) $V_{TH}$ shift during off-state stress at $V_{G,Stress}$ = -5 V.

The $g_m$-$V_G$ curves at $V_D$ = 10 V of the devices after the critical stress steps are shown in Fig. 3 (b)-(d). Maximum drain current ($I_{DSS}$) (at $V_{GS}$ = 0 V, $V_{DS}$ = 15 V), ON resistance ($R_{ON}$) (at $V_{GS}$ = 0 V) and threshold voltage ($V_{TH}$) (at $V_{DS}$ = 10 V) evolution during stress are shown in Fig. 4. It is worth noticing that in $L_{GD}$ = 0.5 µm devices the parametric degradation is less than 10%, until sudden breakdown, when the drain stress voltage is over 65 V.

$I_{DSS}$ decrease and $R_{ON}$ increase started from 85 V and 90 V, for $L_{GD}$ = 1.5 µm and 2.5 µm devices, respectively. The $I_{DSS}$ decrease and $V_{TH}$ shift can be explained by trapping effects under the gate, possibly under the channel or in the buffer [9].

Electroluminescence images of the device with $L_{GD}$ = 0.5 µm ,1.5 µm and 2.5 µm taken before stress and after stress at ($V_G$ = -5 V, $V_D$ = 45 V, 60 V and 70 V) are shown in Fig. 5. Before stress, there is no hot spots along the finger, after 2 minutes' stress at 45 V, 60 V and 70 V, EL image showed leakage points along the gate finger, thus demonstrating the generation of localized leakage paths during stress, possibly in correspondence to defects which may also act as trapping centers.

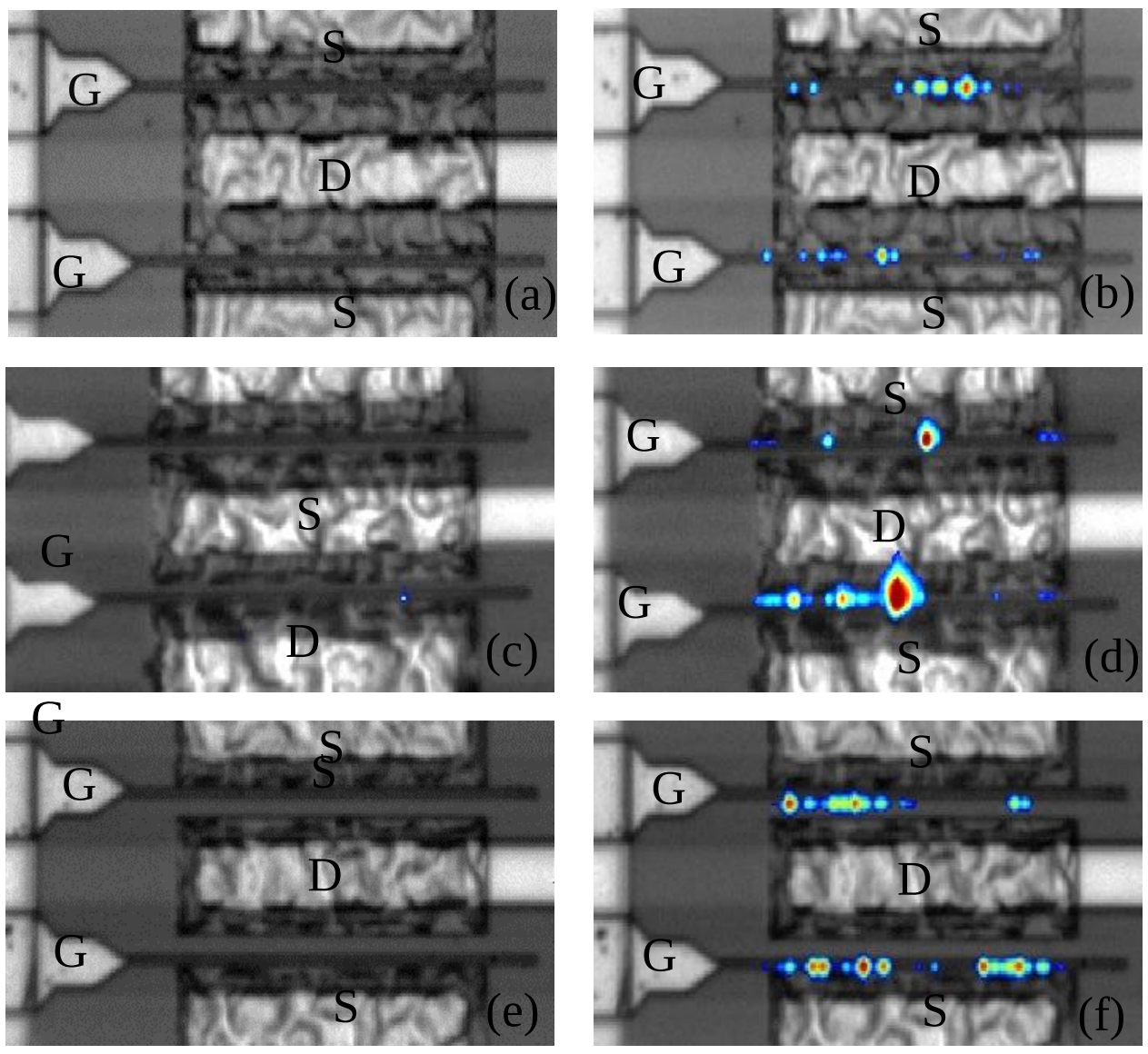

Fig. 5 Emission images (15 s emission) of the $L_{GD}$ = 0.5 µm device (a) before stress at ($V_G$ = -5 V, $V_D$ = 0 V), and (b) during stress at ($V_G$ = -5 V, $V_D$ = 45 V); the $L_{GD}$ = 1.5 µm device (c) before stress at ($V_G$ = -5 V, $V_D$ = 0 V), and (d) during stress at ($V_G$ = -5 V, $V_D$ = 60 V) and the $L_{GD}$ = 2.5 µm device (e) before stress at ($V_G$ = -5 V, $V_D$ = 0 V), and (f) during stress at ($V_G$ = -5 V, $V_D$ = 70 V).

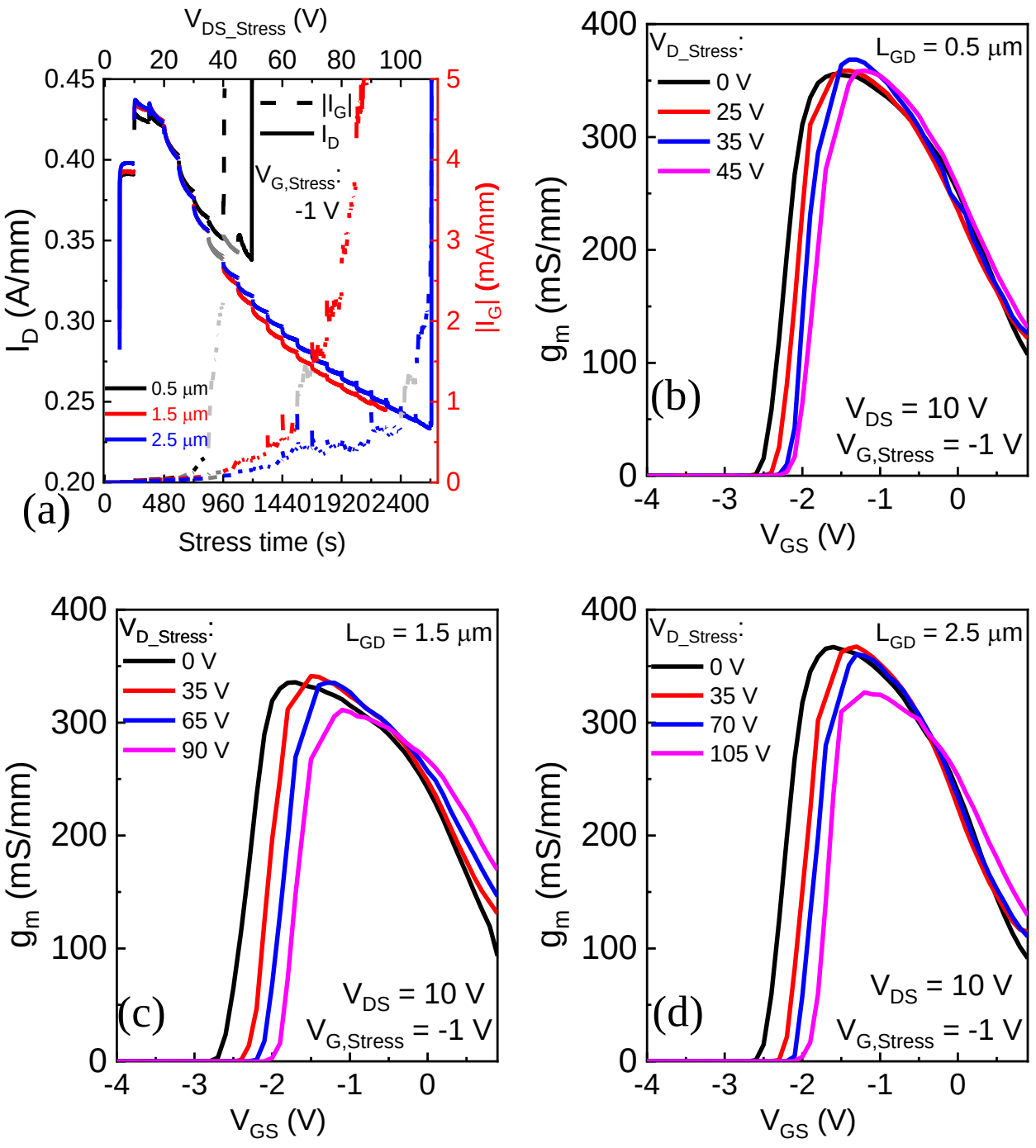

Fig. 6 (a) Drain current monitored during semi on-state stress and (b) $g_m$-$V_G$ curves during stress (b) 0.5 µm (c) 1.5 µm and (d) 2.5 µm devices.

## *B. Semi-on state drain step stress*

Drain and gate currents of the devices with different $L_{GD}$ recorded during semi on-state stress ($V_G$ = -1 V) are shown in Fig.6 Devices with 0.5 µm, 1.5 µm and 2.5 µm $L_{GD}$ showed catastrophic failure at 50 V, 95 V and 110 V, respectively. Channel current decreases during the step stress tests, possibly due to the combined effect of device self-heating and electron trapping. The gate leakage current reached 1 mA/mm at stress voltages of 35 V, 65 V and 100 V, for 0.5 µm, 1.5 µm and 2.5 µm devices, respectively.

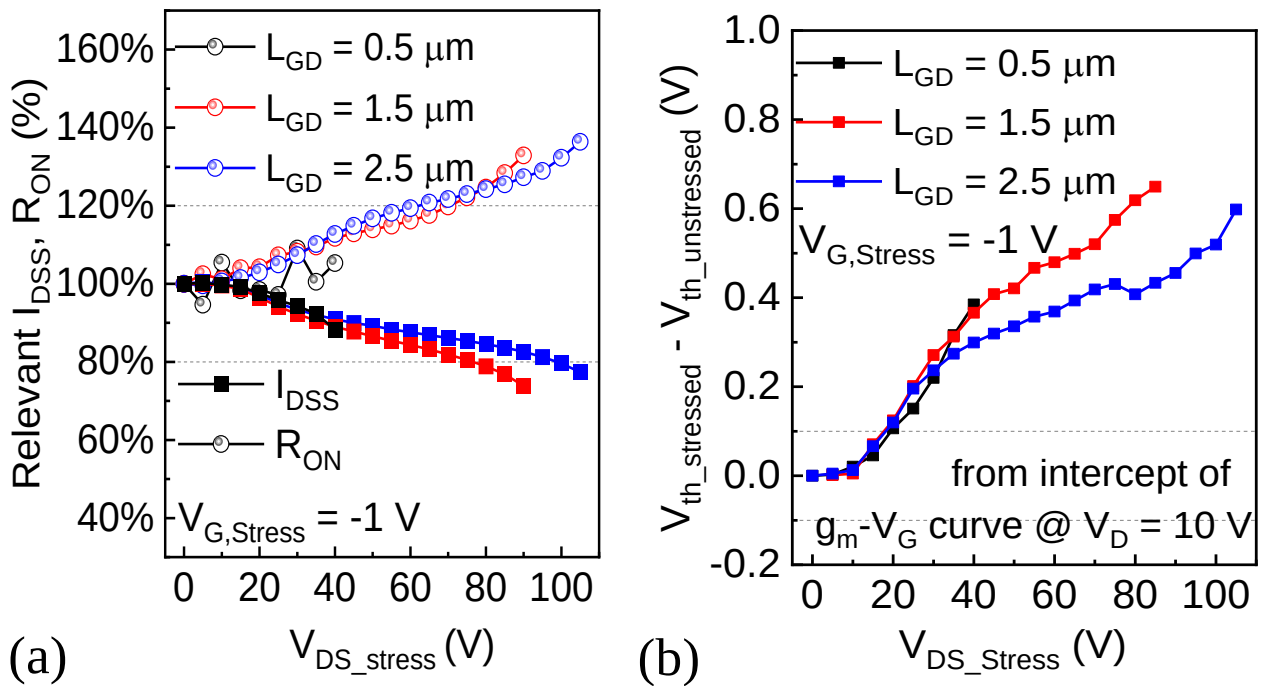

Fig7 (a) $I_{DSS}$, $R_{ON}$ and (b) $V_{TH}$ shift during semi-on stress at $V_{G,Stress}$ = -1 V.

The transconductance curves of the devices at $V_D$ = 10 V after some stress steps are shown in Fig. 6 (b)-(d). Different from what was observed during off-state stress, $I_{DSS}$ decreased gradually, $R_{ON}$ increased gradually, and $V_{TH}$ showed gradual positive shift with increasing stress voltage in all devices, regardless of $L_{GD}$, as shown in Fig. 7. The $L_{GD}$ = 2.5 µm device showed the largest maximum transconductance decrease at the step before breakdown.

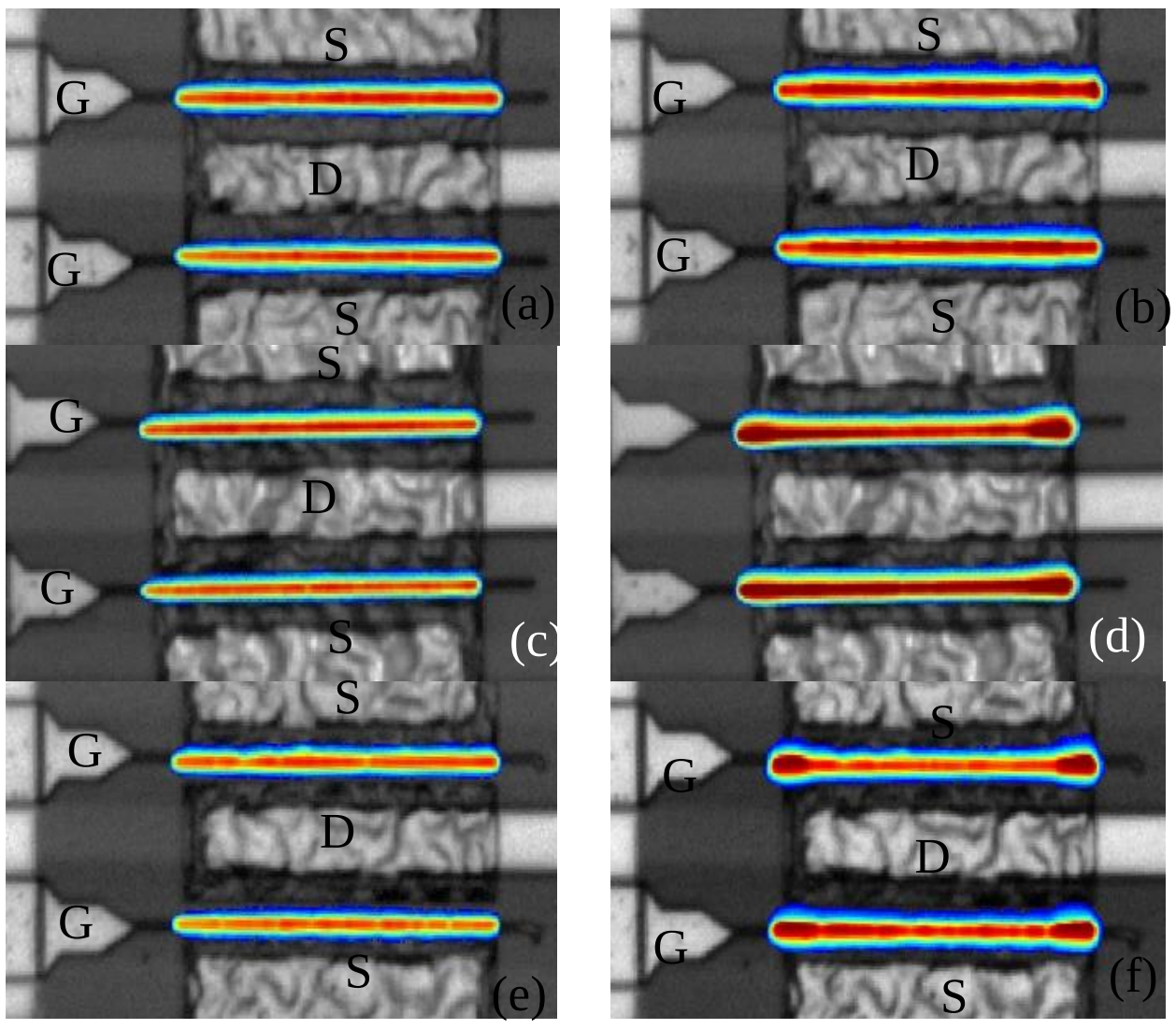

Fig. 8 Emission images (15 s emission) of the $L_{GD}$ = 0.5 µm device (a) at ($V_G$ = -1 V, $V_D$ = 20 V), and (b) during stress at ($V_G$ = -1 V, $V_D$ = 30 V); the $L_{GD}$ = 1.5 µm device (c) at ($V_G$ = -1 V, $V_D$ = 20 V), and (d) during stress at ($V_G$ = -1 V, $V_D$ = 40 V) and the $L_{GD}$ = 2.5 µm device (e) at ($V_G$ = -1 V, $V_D$ = 20 V), and (f) during stress at ($V_G$ = -1 V, $V_D$ = 40 V).

No leakage current increase was observed until the stress voltage exceeded 25 V, 35 V and 35 V for 0.5 µm, 1.5 µm and 2.5 µm devices, respectively, in agreement with the gate leakage current change observed during stress.

The EL images of the devices with $L_{GD}$ = 0.5 µm, 1.5 µm and 2.5 µm taken during stress at stress ($V_D$ = 20 V) and after stress at ($V_D$ = 40 V) are shown in Fig. 8. Even EL intensity distribution can be observed along gate fingers during tests at $V_D$ = 20 V, as shown in Fig. 8 (a) (c) (e). At higher drain voltages, emissions are stronger at finger extremes, as carrier transport is worsened by device self-heating in the central part of the device. Another explanation could be that the peak electric field at the drain edge of the gate-drain access region increases with increasing drain stress voltage, boosting carrier conduction, causing increased emission intensity [10].

## C. *On-state drain step stress*

Drain and gate current of the devices recorded during on-state stress are shown in Fig. 6 (a). Despite the very large power density (15 W/mm ∽ 30 W/mm), devices with 0.5 μm, 1.5 μm and 2.5 μm of $L_{GD}$ did not fail up to burnout occurring at 30 V, 55 V and 65 V, respectively.

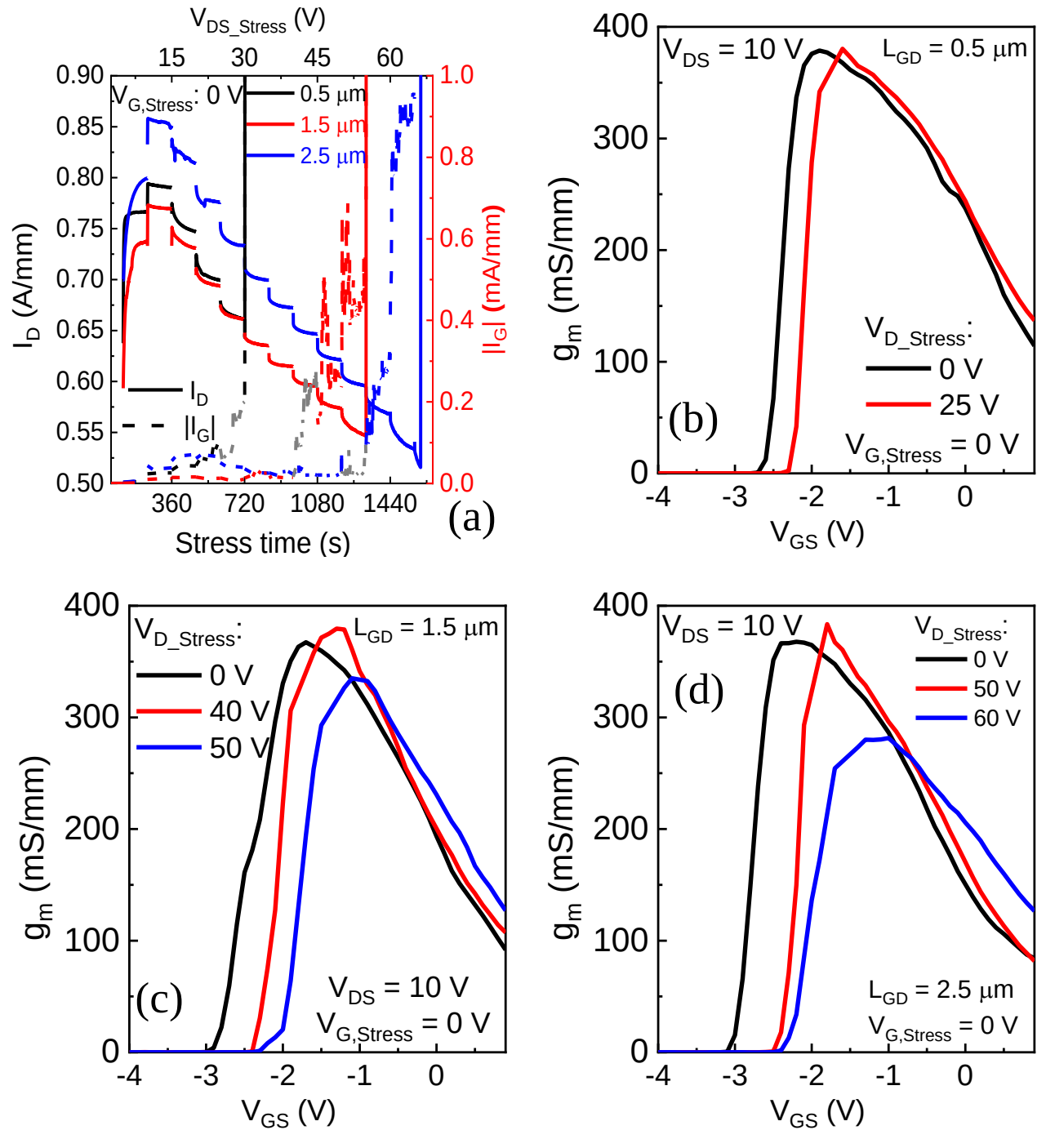

Fig. 6 (a) Drain current monitored during on-state stress, $g_m$-$V_G$ curves during stress (b) 0.5 μm (c) 1.5 μm and (d) 2.5 μm devices.

The $g_m$-$V_G$ curves of the devices at $V_D$ = 10 V after several stress steps are shown in Fig. 6(b)-(d). Similar to the semi-on state stress, $g_{m,max}$ decrease occurred at the last stress step before breakdown, and the devices with $L_{GD}$ = 2.5 μm have showed maximum $g_{m,max}$ degradation.

$V_{TH}$ shifts positively with increasing stress voltage, with comparable shift in devices with different $L_{GD}$,, possibly due to traps under the gate in the buffer or at gate edges, whose occupation and effect does not depend on the gate-drain distance, as shown in Fig. 7.

Leakage current increase can be observed, at 25 V, 40 V and 50 V for 0.5 μm, 1.5 μm and 2.5 μm devices, consistent with the gate leakage current increase monitored during stress.

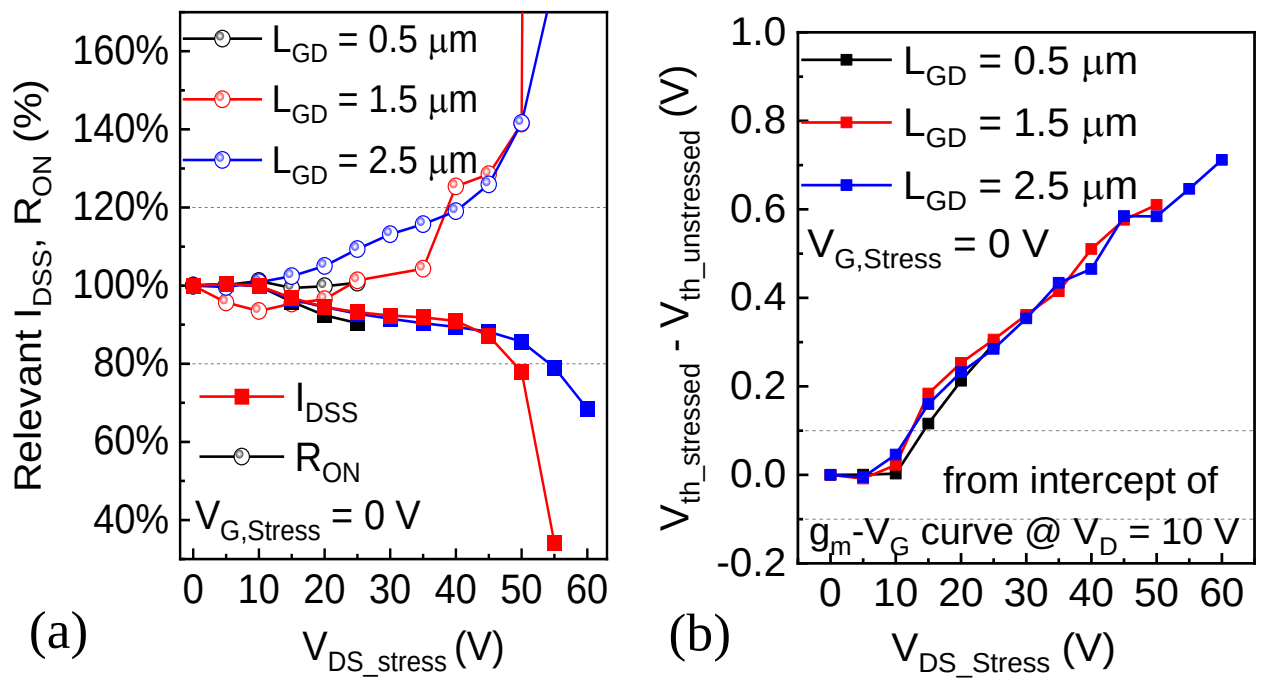

Fig. 7 (a) $I_{DSS}$, $R_{ON}$ and (b) $V_{TH}$ shift during on-state stress at $V_{G,Stress}$ = 0 V.

Fig. 8 shows breakdown voltages $V_{BR}$ and critical voltage $V_{CRI}$ (The voltage where leakage current exceeds 1 mA/mm) of the AlN/GaN HEMTs with difference $L_{GD}$. DC short-term safe operating area for the three different gate-drain spacings are also shown. It should be noted that, despite the thin barrier layer (3 nm AlN) and short channel length (110 nm), devices with the shortest $L_{GD}$ reach off-state $V_{BR}$ = 70 V, $V_{CRI}$ = 55 V (50 V and 35 V in semi-on, on state) and can withstand a DC dissipated power as high as $P_D$ = 15 W/mm in on-state.

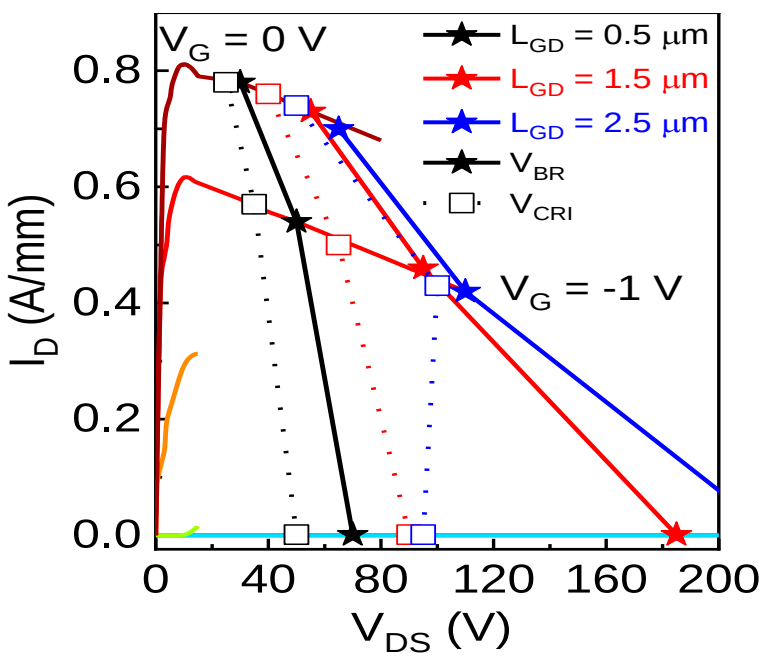

Fig. 8 $V_{BR}$ comparison among devices with different $L_{GD}$.

## D. *Constant Voltage stress- off state*

In order to evaluate the reliability of the devices under operating conditions, constant voltage stress tests in the off-state condition $V_G$ = -5 V, $V_D$ = 40 V were carried out. DC characteristics were measured at room temperature before, during (at each step) and after the stress.

Drain and gate leakage current during stress are shown in Fig. 11. Leakage current reached 1 mA/mm after four hours' stress for the device with 0.5 μm of $L_{GD}$, possibly due to the leakage path caused by trapping effects assisted by high electric field at the gate edge. The device with 1.5 μm and 2.5 μm of $L_{GD}$ showed less than one order of magnitude increase in leakage current after 24 hours' stress.

The $I_D$-$V_G$ curves of the devices at $V_D$ = 10 V before and after 24 hours' stress are shown in Fig. 11(b)-(d). The $L_{GD}$ = 0.5 μm devices showed one order of magnitude increase in leakage current, a value in agreement with the leakage current increase observed during stress. The threshold voltage shifted positively by +0.3 V. The $g_m$-$V_G$ curves in Fig. 9(a) showed that there is no $g_{m,max}$ decrease during stress.

For the devices with $L_{GD}$ = 1.5 μm, there is almost no leakage current increase, similar to that observed during stress. The transfer characterization $I_DV_G$ in Fig. 11(c) and $g_m$-$V_G$ curves in Fig. 9 (b) showed that the degradation signature of the devices is $V_{th}$ positive shift (+0.2 V). For the devices with $L_{GD}$ = 2.5 μm, no leakage current increase or maximum transconductance increase is observed, similar to the device with $L_{GD}$ = 1.5 μm, and the $V_{TH}$ shift is close to +0.18 V.

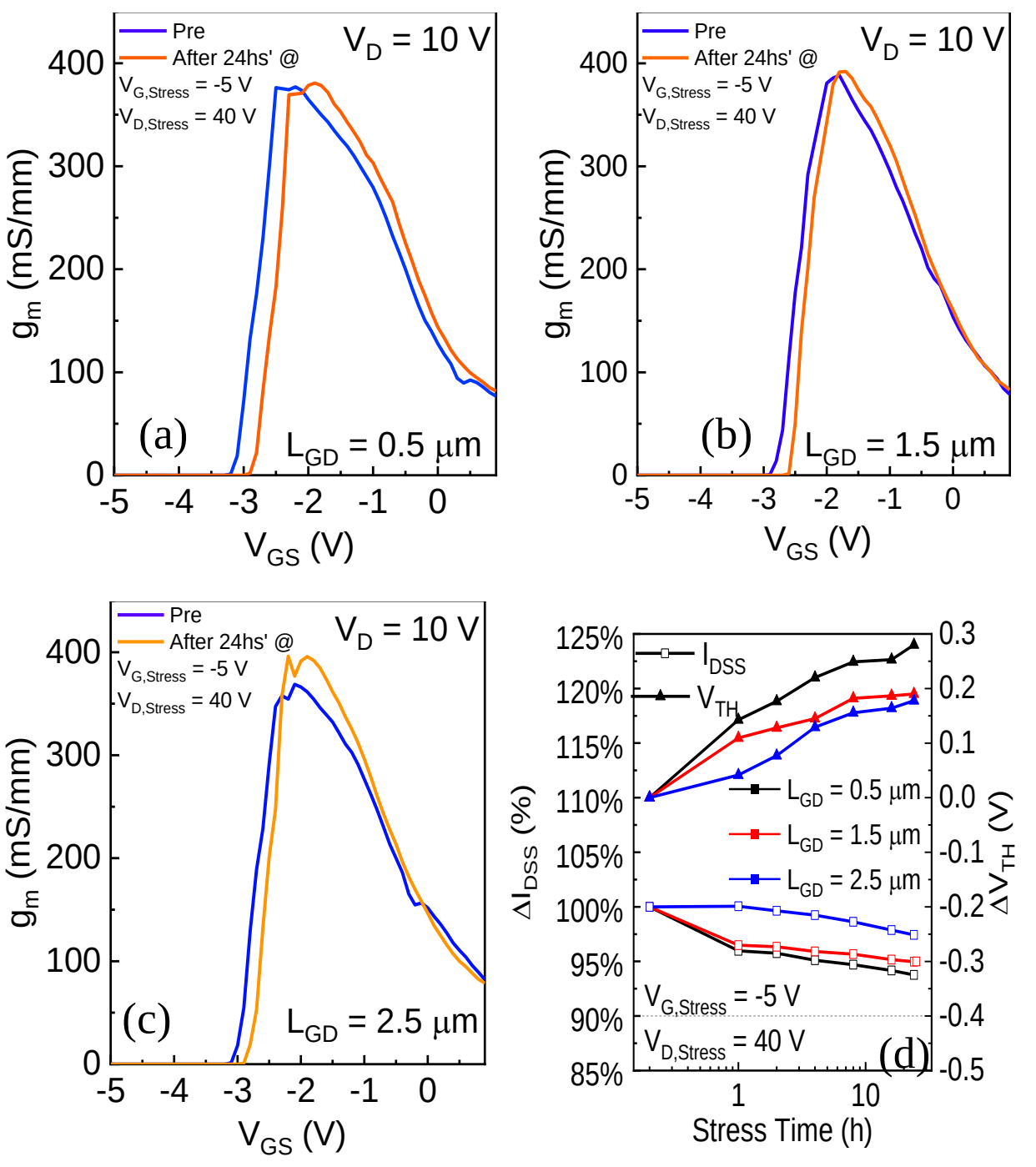

Fig. 9 Transfer $g_m$-$V_G$ curves during off-state stress of (a) 0.5 μm (b) 1.5 μm and (c) 2.5 μm devices and (d) $I_{DSS}$ and $V_{TH}$ shift during 24 hours constant voltage stress at $V_{G,Stress}$ = -5 V, $V_{D,Stress}$ = 40 V.

The $I_{DSS}$ decrease and $V_{TH}$ shift during stress are summarized in Fig. 9 (d). The $I_{DSS}$ decrease is less than 10%, and $I_{DSS}$ degradation percentage and $V_{TH}$ shift showed a dependence on $L_{GD}$.

Electroluminescence images of the three kinds of devices taken during the first 15 s and after 24 hours stress at ($V_G$ = -5 V, $V_D$ = 40 V) are shown in Fig. 10. Five hot spots occurred as soon as stress started for the 0.5 μm device, three hot spots can be observed for the 1.5 μm device, and one hot spot can be observed for the 2.5 μm device. After stress at (-5 V, 40 V) for 24 hours, EL image showed many new leakage points along the gate finger. Therefore, this demonstrates that the generation of localized leakage paths during stress, possibly in correspondence to defects which may also act as trapping centers.

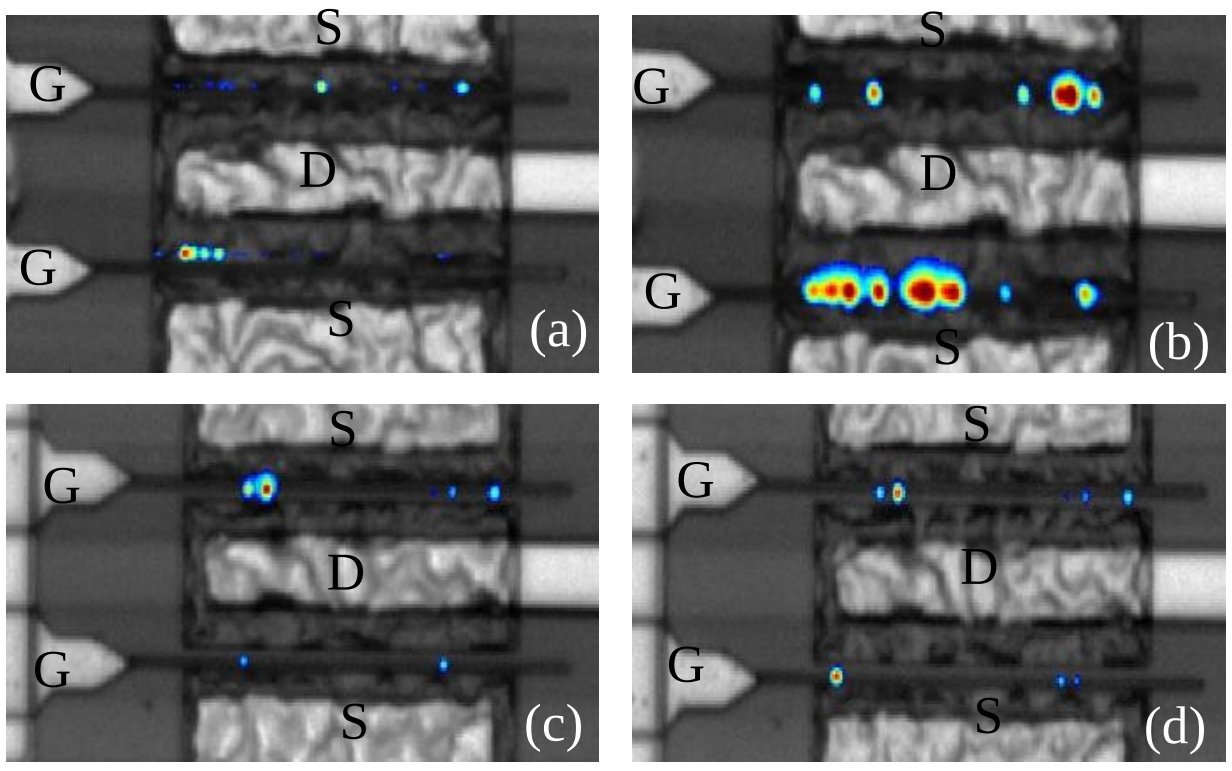

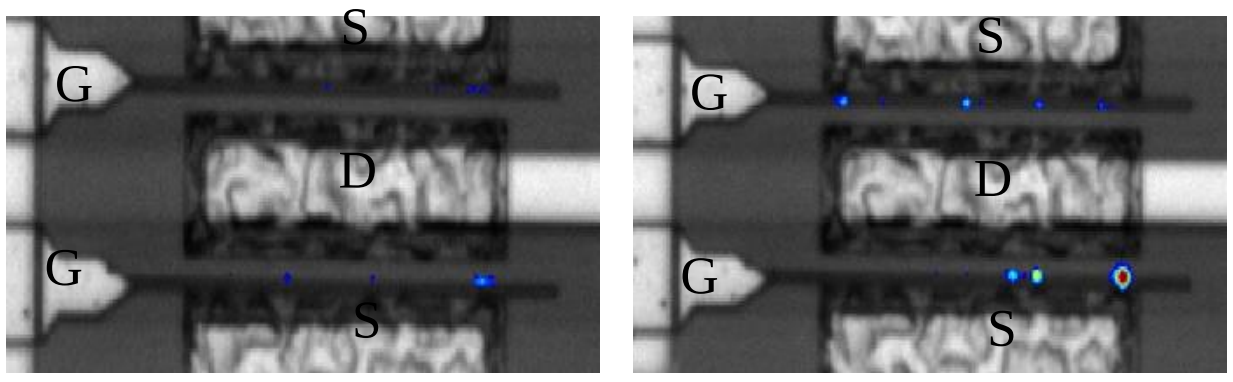

Fig. 10 Emission images (15 s) of the (a) $L_{GD}$ = 0.5 μm device at the first 15 s stress at (-5 V, 40 V) (b) after 24 hours stress at (-5 V, 40 V) 1.5 μm and 2.5 μm devices.

From the results shown above, positive threshold voltage shift and leakage current increase has been observed during all the stress processes, regardless of bias voltage and power disspation (off state, semi-on state and on state). Maximum transconductance decrease was observed at the step before breakdown during semi-on and on state stress, and the degradation percentage depends greatly on $L_{GD}$, thus suggesting increase of parasitic drain resistance and worsening of carrier transport properties in the access region.

The $V_{TH}$ shift during the off-state stress could be explained by trapping effects under the gate, traps could be generated in the GaN buffer far below the 2DEG [11], result from vacancy migration under high electric field [12], or can be related to the carbon doping in the GaN buffer [13].

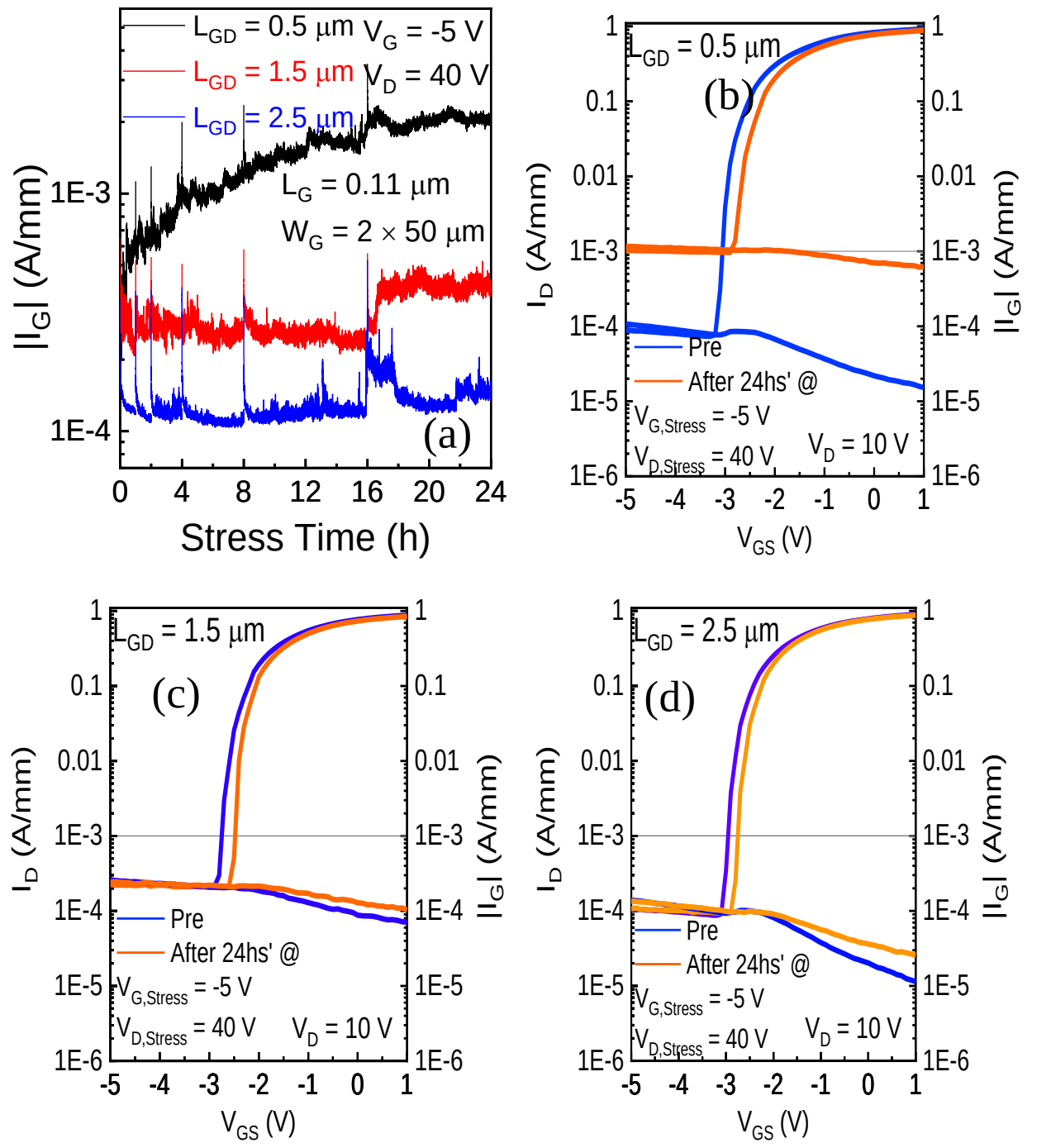

Fig. 11 (a) Drain current monitored during 24 hours constant voltage stress in off-state, $I_D$-$V_G$ curves during stress (b) 0.5 μm (c) 1.5 μm and (d) 2.5 μm devices.

Another explanation could be that during off state stress, electron tunneling occurs at the drain edge of the gate, and electrons are injected into the surface-state layer and captured by the surface deep donors, or charges at the SiN/AlN interface [14], hereby, the 2DEG density is decreased due to these negative surface charges, leading to $V_{TH}$ positive shift [15].

## IV. CONCLUSIONS

In conclusion, short-term reliability and robustness of 110 nm AlN/GaN HEMTs has been evaluated by means of off-state, semi-on state and on-state step stress tests and a constant voltage stress at off-state on devices having different gate-drain distance, $L_{GD}$. Results proved that the new short channel, ultra-thin barrier devices have an excellent device robustness. The breakdown voltages and critical voltages scale almost linearly with $L_{GD}$, failure mode remains almost unchanged in the nine device groups, and consists in an increase of gate leakage, accompanied by a positive shift of threshold voltage. In off-state, electroluminescence images detect the presence of localized leakage paths which may correspond to dislocations and act as preferential paths for electron trapping. Failure modes do not depend on power dissipation and, as a consequence, on temperature. Degradation is therefore preliminary attributed to trapping effects, enhanced by electric field, as shown in [14], [15].

## ACKNOWLEDGEMENT

Support by EUGANIC project under the EDA Contract B 1447 IAP1 GP, by the EC Horizon 2020 ECSEL project 5G_GaN_2, by the ESA ESTEC project RELGAN, and by the Italian Ministry of University and Research (MIUR), PRIN project GANAPP is gratefully acknowledged.